\documentclass[reprint,twocolumn,superscriptaddress,showpacs,preprintnumbers,nofootinbib,aps,prd,
]{revtex4-1}

\usepackage{graphicx,color}
\usepackage{amsmath,amssymb,amsfonts}
\usepackage{stackrel}
\usepackage{enumitem}
\usepackage[english]{babel}
\usepackage{verbatim}
\usepackage{hyperref}
\usepackage{comment}
\usepackage{ulem}
\usepackage{xcolor}
\usepackage{subcaption}
\usepackage{dcolumn}
\usepackage{physics}

\graphicspath{
  {./}
  {figures/}
}

\newcommand{\tf}{\texorpdfstring}
\newcommand{\gev}{~\text{GeV}}
\newcommand{\tev}{~\text{TeV}}
\newcommand{\mev}{~\text{MeV}}

\newcommand{\onbb}{0\nu\beta\beta}

\def\nn{\nonumber}
\newcommand{\slashed}{\slash \hspace{-0.23cm}}

\newcommand{\abi}{~\text{ab}^{-1}}
\newcommand{\fbi}{~\text{fb}^{-1}}

\definecolor{orange}{rgb}{1,0.5,0}
\definecolor{amethyst}{rgb}{0.6, 0.4, 0.8}
\definecolor{antiquefuchsia}{rgb}{0.57, 0.36, 0.51}
\definecolor{byzantine}{rgb}{0.74, 0.2, 0.64}
\definecolor{blue-violet}{rgb}{0.54, 0.17, 0.89}
\definecolor{cadmiumred}{rgb}{0.89, 0.0, 0.13}
\definecolor{brightcerulean}{rgb}{0.11, 0.67, 0.84}

\begin{abstract}
\noindent
We study the complementary tests of lepton number violation in $0\nu\beta\beta$-decay experiments, long-lived particle (LLP) searches at the LHC main detectors ATLAS/CMS, and a proposed far detector MATHUSLA. In the context of a simplified model with a scalar doublet $S$ and a Majorana fermion $F$, we show that while the $\onbb$-decay experiments can probe a larger portion of parameter space,  the LLP searches can uniquely probe the region of smaller couplings and masses if $S$ is at TeV scale while $F$ is at or below the electroweak scale. We also investigate constraints on the parameter space from the existing searches that are insensitive to lepton number violation.
\end{abstract}

\begin{document}

%% Setting the space between equations
\setlength{\abovedisplayskip}{6pt}
\setlength{\belowdisplayskip}{6pt}
%% addresses

\newcommand{\SJTU}{\affiliation{Tsung-Dao Lee Institute and School of
    Physics and Astronomy, Shanghai Jiao Tong University, 800
    Dongchuan Road, Shanghai, 200240 China }}

\newcommand{\UMASS}{\affiliation{Amherst Center for Fundamental
    Interactions, Department of Physics, University of Massachusetts,
    Amherst, MA 01003} }

\newcommand{\CIT}{\affiliation{Kellogg Radiation Laboratory,
    California Institute of Technology, Pasadena, CA 91125 USA} }
    
\newcommand{\UA}{\affiliation{Department of Physics, University of Arizona, Tucson, Arizona 85721, USA}}

\title{Lepton Number Violation: from $0\nu\beta\beta$ Decay to Long-Lived Particle Searches}

\preprint{ACFI-T21-11}

\author{Gang~Li}
\email{ligang@umass.edu}
\UMASS

\author{Michael~J.~Ramsey-Musolf}
\email{mjrm@sjtu.edu.cn, mjrm@physics.umass.edu}
\SJTU\UMASS\CIT

\author{Shufang~Su}
\email{shufang@arizona.edu}
\UA

\author{Juan~Carlos~Vasquez}
\email{jvasquezcarm@umass.edu}
\UMASS

\maketitle

\section{Introduction}
In the Standard Model of particle physics, total lepton (L) and baryon (B) numbers are conserved at the classical (Lagrangian) level, while the $B+L$ anomaly breaks this conservation law through quantum corrections. It is possible that physics beyond the Standard Model (BSM) introduces B and/or L-violating interactions in the Lagrangian. The well-known seesaw mechanism~\cite{Minkowski:1977sc,Mohapatra:1979ia,Glashow:1979nm,Gell-Mann:1979vob,Schechter:1980gr,Yanagida:1980xy}
for neutrino mass provides a strong motivation for the existence of low-energy lepton number violating (LNV) interactions, such as the dimension-5  $\Delta L=2$ \lq\lq Weinberg operator"~\cite{Weinberg:1979sa} that implies the existence of a light neutrino Majorana mass.

If LNV interactions exist in Nature, exploring the possible associated mass scale(s) $\Lambda$ is interesting. For $\Lambda$ of order the conventional seesaw scale $\sim 10^{12}\gev-10^{15}\gev$, direct observation of the responsible BSM particles and interactions in laboratory or astrophysical signatures is unlikely. Instead, the most experimentally accessible effects are associated with the non-renormalizable LNV effective operators involving only light degrees of freedom, such as the Weinberg operator mentioned above. 

It is entirely plausible, however, that $\Lambda$ lies well below the seesaw scale. In this work, we consider experimental signatures when $\Lambda$ is of $\mathcal{O}(\mathrm{TeV})$ and below, making it at least in principle possible to observe the BSM degrees of freedom directly. Theoretically, TeV scale lepton number violation arises in various well-studied models, such as the
R-parity violating supersymmetry (SUSY)~\cite{Dreiner:1997uz,Allanach:2003eb,Barbier:2004ez} 
and the minimal Left-Right Symmetric Model (mLRSM)~\cite{Pati:1974yy,Mohapatra:1974gc,Senjanovic:1975rk,Mohapatra:1979ia,Mohapatra:1980yp,Senjanovic:1978ev}.  Experimentally, the most promising signatures include the neutrinoless double beta-decay ($\onbb$-decay) of atomic nuclei and LNV processes in high-energy proton-proton collisions.

Numerous studies have considered the implications of TeV-scale lepton number violation for $\onbb$-decay~\cite{Hirsch:1995zi,Hirsch:1995cg,Hirsch:1996qw,Chakrabortty:2012mh,Barry:2013xxa,BhupalDev:2014qbx,Helo:2015fba} and the connection with searches at high-energy colliders~\cite{Datta:1993nm,Allanach:2009iv,Tello:2010am,Nemevsek:2011aa,Das:2012ii,Helo:2013dla,Helo:2013ika,Chen:2013foz,Peng:2015haa,Deppisch:2015qwa,Gonzalez:2016ztm,Lindner:2016lpp,Lindner:2016lxq,Cepedello:2017lyo,Cai:2017mow,Nemevsek:2018bbt,Harz:2021psp,Chauhan:2021xus}.
In what follows, we focus on the interplay of these two classes of signatures, with a particular emphasis on the possibility that one or more of the BSM particles may be relatively long-lived. The search for long-lived particles (LLPs) at the Large Hadron Collider (LHC), as well as prospective future $e^+e^-$ and $pp$ colliders, has received considerable recent attention (for a review of the LHC prospects, see Ref.~\cite{Alimena:2019zri}).   

To investigate the possible complementarity between $\onbb$-decay and LLP searches, we adopt a simplified model framework that has been utilized previously to explore the $\onbb$-decay/collider interplay~\cite{Peng:2015haa}. The spirit of adopting the simplified model in our work is to draw possible connections between LLP searches and $\onbb$ decay. The choices of simplification in the model are intended to highlight these connections in a way that has some degree of generality. In this respect, it is useful to observe that -- for purposes of analyzing $\onbb$-decay -- one may map different models onto a finite set of non-renormalizable operators containing only Standard Model (SM) quark and lepton fields. The chiral transformation properties of the corresponding hadronic components then determine, at the level of Weinberg chiral power counting, the expected importance of their contribution to the $\onbb$-decay rate~\cite{Prezeau:2003xn,Graesser:2016bpz,Cirigliano:2017djv,Cirigliano:2018yza}.
The simplified model of Ref.~\cite{Peng:2015haa} induces the leading order (LO) long-range pion-exchange $\onbb$-decay amplitude that one nominally expects to induce the largest impact on the $\onbb$-decay rate. Only a subset of simplified models have this feature, see Refs.~\cite{Prezeau:2003xn,Graesser:2016bpz} and references therein.
Of these, the model that we adopt here extends the SM with a minimal set of particles and interactions. We defer to future work a treatment of the phenomenology of $\onbb$-decay, collider probes, and other experimental tests for simplified models that do not exhibit this LO chiral amplitude and minimality.

In this context of our simplified model, one of the new degrees of freedom may be long-lived when it is weakly coupled and relatively light with its mass $\sim \mathcal{O}(50\, \mathrm{GeV})$. For the LLP searches at the LHC, we consider both ATLAS and CMS capabilities, as well as the proposed MATHUSLA detector~\cite{Curtin:2018mvb}\footnote{We find that the FASER detector~\cite{Feng:2017uoz} would not exhibit sensitivity to the simplified model we consider without invoking strong assumptions about the model flavor structure.}.  We find that $\onbb$-decay, the LHC main detectors, and MATHUSLA provide richly complementary probes of the model parameter space. Notably,  since $\onbb$-decay generally provides the widest sensitivity to LNV interactions involving first-generation SM fermions, its observation would provide no direct information about the underlying particle physics mechanism or  $\Lambda$. Uncovering both the mechanism and LNV mass scale will require additional experimental handles. Our following study illustrates the potential for LLP searches to provide one such handle.

\section{Model and \tf{$\onbb$}{0vbb}-decay }
\label{sec:model_DBD}

We adopt a simplified model for our analyses, which was introduced in Ref.~\cite{Peng:2015haa}. Within this simplified framework,  the interplay between LHC searches with prompt decaying particles and $\onbb$-decay has been discussed in Ref.~\cite{Peng:2015haa}, where it was concluded that ton-scale $\onbb$-decay experiments reach generally exceeds that of the LHC with an integrated luminosity of 300 ${\rm fb}^{-1}$. However, both ton-scale $\onbb$-decay experiments and the searches at the high-luminosity LHC (HL-LHC) are complementary for TeV scale masses. 

In Ref.~\cite{Peng:2015haa} and references therein, it has also been pointed out that for TeV scale LNV interactions, one needs to calculate the $\onbb$-decay rate in the effective field theory (EFT) approach, especially chiral perturbation theory at low energy.
As discussed in Ref.~\cite{Prezeau:2003xn},  several quark-lepton effective operators  can give rise to leading-order (LO) $\pi\pi ee$ interactions in the chiral Lagrangian,   which can hence dominate the $\onbb$-decay rate~\footnote{For recent EFT analyses, see, {\it e.g.} Refs.~\cite{Graesser:2016bpz,Cirigliano:2017djv,Cirigliano:2018yza}}. 
Such effective operators can be obtained by integrating out heavy fields in ultraviolet (UV) theories, such as the RPV SUSY and the mLRSM. One possible minimal model that gives rise to the LO $\pi\pi ee$ interaction includes a scalar SM SU(2) doublet $S\in (1,2)_{1/2}$  and a Majorana fermion singlet $F\in (1,1)_0$,  where $(X,Y)_Z$ denotes transformation properties under the SM gauge group~\cite{Peng:2015haa}.  The Lagrangian of this model  can be written as \begin{align}
\label{eq:Lag}
\mathcal{L}&= (\partial_\mu S)^\dagger \partial^\mu S - m_S^2 S^\dagger S + \dfrac{1}{2}\bar{F}^c (i\slashed{\partial}-m_F)F \nn\\
&\quad + g_Q \bar{Q}_L S d_R + g_L \bar{L} \tilde{S} F  
+ \text{h.c.}\;,\;
\end{align}
alongside with the SM interactions. Here, $\tilde{S}\equiv i\tau^2 S^*$, $\tau^i, i=1,2,3$ are the Pauli matrices, $Q=(u,d)_L^T$ and $L=(\nu,e)_L^T$ are the weak isospin doublets of the first-generation left-handed quarks and leptons, and ``h.c.'' denotes the Hermitian conjugation. In the full theory such as the RPV SUSY~\cite{Allanach:2003eb}, $S$ and $F$ are identified as the slepton and the lightest neutralino fields, respectively.  
Without loss of generality, we will assume $g_L$ and $g_Q$ are real and positive. Notice that other terms are possibly allowed, such as the term $\lambda_{HS}(H^\dagger S)^2$ that induces neutrino mass at one-loop level~\cite{Harz:2021psp}, the size of which is proportional to the scalar coupling $\lambda_{HS}$. For simplicity, we will   omit such allowed terms 
and assume that $S$ does not develop a non-zero  vacuum expectation value (VEV) $\langle S \rangle$.  Note that if there is sizable $S-H$ mixing or if  $S$ develops a non-zero VEV that is not small,  the size of $g_L$ in Eq.~\eqref{eq:Lag} could be highly suppressed by the smallness of neutrino mass.  The phenomenology described in this paper could be significantly changed or even disappear.  In our analyses below, we focus on the case when $S$ is inert and does not develop a sizable VEV.

In our work, we will consider an alternative scenario with a light Majorana fermion $F$, that could potentially lead to a LLP signature at colliders~\footnote{It is interesting to notice that a light $F$ with the mass as low as $500\mev$ can satisfy all the cosmological constraints in the context of RPV SUSY, see Refs.~\cite{Dreiner:2020qbi,Dreiner:2009ic} and reference therein. 
}.  If $S$ is at the TeV scale while $F$ is much lighter, these fields are integrated out separately when deriving the low-energy EFT~\footnote{We will assume $m_F\geq 2\gev$. For $m_F< 2\gev$, $F$ is  not integrated out in the effective field theory approach to $\onbb$-decay as studied in Ref.~\cite{Dekens:2020ttz}.}. The scalar $S$ is firstly integrated out at the scale $\mu=m_S$ and the following dimension-6 (dim-6)  lepton-number-conserving effective Lagrangian is obtained~\cite{Dekens:2020ttz}
\begin{align}
\label{eq:dim-6_operator}
\mathcal{L}_{\Delta L = 0}^{(6)}&=\dfrac{2G_F}{\sqrt{2}} \Big[  C_{\text{SRR}}^{(6)} \bar{u}_L d_R \bar{e}_L  F_R  + \text{h.c.}\Big]\;,
\end{align}
where $F_R\equiv P_R F$,  $P_R = (1+\gamma_5)/2$, and $G_F$ is the Fermi constant.  The non-vanishing Wilson coefficient $C_{\text{SRR}}^{(6)}(m_S)$  is given by  $C_{\text{SRR}}^{(6)}(m_S) = g_L g_Q v^2/m_S^2$ with $v=246\gev$. The renormalization group equation (RGE) of $C_{\text{SRR}}^{(6)}$ under QCD running from $m_S$ to $m_F$ is described as $d C_{\text{SRR}}^{(6)}/d\ln \mu=-2\alpha_S/\pi C_{\text{SRR}}^{(6)}$~\cite{Dekens:2020ttz,Jenkins:2017dyc,Aebischer:2017gaw}, where $\alpha_S$ is the running strong coupling. For $m_F=30\gev$ and $m_S=1\tev$,  $C_{\text{SRR}}^{(6)}(m_F)=1.28\ C_{\text{SRR}}^{(6)}(m_S)$.

The Majorana fermion field $F$ is integrated out at $\mu=m_{F}$ and the dim-6 effective Lagrangian in Eq.~\eqref{eq:dim-6_operator} is matched to the following dim-9 Lagrangian~\cite{Prezeau:2003xn,Cirigliano:2018yza},
\begin{align}
\mathcal{L}_{\Delta L = 2}^{(9)} \supset \dfrac{1}{v^5} C_{2L}^{(9)\prime} O_2^\prime  \bar{e}_L  e_L^c + \text{h.c.}\;,
\end{align}
where $O_2^\prime \equiv (\mathcal{O}_{2+}^{++}-\mathcal{O}_{2-}^{++})/2$  and $e_L^c \equiv (e_L)^c$. Here, the operators $\mathcal{O}_{2\pm}^{++}$ were introduced in Ref.~\cite{Prezeau:2003xn}, which are 
\begin{align}
\mathcal{O}_{2\pm}^{++}=(\bar{q}_R \tau^+ q_L)(\bar{q}_R \tau^+ q_L)\pm (\bar{q}_L \tau^+ q_R)(\bar{q}_L \tau^+ q_R)\;,
\end{align}
where $q_{L,R}=(u,d)^T_{L,R}$ are the left-handed and right-handed quark isospin doublets, $\tau^+=(\tau^1 + i\tau^2)/2$. 

The matching condition at $\mu=m_{F}$ is:
\begin{align}
C_{2L}^{(9)\prime}(m_F) &=\dfrac{v}{2} \left( C_{\text{SRR}}^{(6)} \right)^2 m_F^{-1}\;.
\end{align}

Below $\mu=m_F$, the operator $O_2^\prime\bar{e}_L e_L^c$ evolves under QCD running and mix with the color-mixed operator $O_3^\prime\bar{e}_L e_L^c$, which is defined as~\cite{Cirigliano:2018yza,Graesser:2016bpz}
\begin{align}
O_3^\prime&=(\bar{q}_L^\alpha \tau^+ q_R^\beta)(\bar{q}_L^\alpha \tau^+ q_R^\beta)\;
\end{align}
with $\alpha,\beta$ being the color indices. The RGEs of the corresponding Wilson coefficients $C_{2L}^{(9)\prime}$ and $C_{3L}^{(9)\prime}$, which is defined as $\mathcal{L}_{\Delta L = 2}^{(9)} \supset 1/v^5 C_{3L}^{(9)\prime} O_3^\prime  \bar{e}_L  e_L^c + \text{h.c.}$, are described as~\cite{Cirigliano:2018yza,Liao:2019gex,Peng:2015haa}
\begin{align}
\dfrac{d}{d\ln \mu} 
\begin{pmatrix}
C_{2L}^{(9)\prime}\\
C_{3L}^{(9)\prime}
\end{pmatrix}
=\dfrac{3\alpha_s}{16\pi}
\begin{pmatrix}
-7 & 4\\
1 & 8
\end{pmatrix}
\begin{pmatrix}
C_{2L}^{(9)\prime}\\
C_{3L}^{(9)\prime}
\end{pmatrix}\;,
\end{align}
For $m_F=30\gev$, we obtain that $C_{2L}^{(9)\prime}(m_0)=1.56\ C_{2L}^{(9)\prime}(m_F)$ and $C_{3L}^{(9)\prime}(m_0)=-0.06\ C_{2L}^{(9)\prime}(m_F)$ with $m_0=2\gev$.

Below $\mu=\Lambda_\chi$ with $\Lambda_\chi\sim 1\gev$ being the chiral symmetry breaking scale, the quark-lepton operators $O_{2,3}^{\prime}\bar{e}_L e_L^c$   are mapped onto the hadron-lepton operators using chiral perturbation theory~\cite{Prezeau:2003xn,Cirigliano:2018yza}, leading to
\begin{align}
\mathcal{L}_{\pi\pi} = -\dfrac{F_\pi^2}{2v^5} C_{\pi\pi L}^{(9)} \pi^-\pi^- \bar{e}_L e_L^c+\text{h.c.}\;,
\end{align}
where $F_\pi=92.4\mev$ is the pion decay constant and the coefficient
\begin{align}
&C_{\pi\pi L}^{(9)} = g_2^{\pi\pi} C_{2L}^{(9)\prime}  + g_3^{\pi\pi}  C_{3L}^{(9)\prime} \;, 
\end{align}
 with the low energy constants  $g_2^{\pi\pi}=(2.0\pm 0.2)\gev^2$ and $g_3^{\pi\pi}=-(0.62\pm 0.06)\gev^2$~\cite{Cirigliano:2018yza,Nicholson:2018mwc}. 
In principle, there are other contributions involving a one-pion exchange or four-nucleon contact interactions, which are sub-leading in chiral power counting, which we neglect here~\cite{Prezeau:2003xn}.  

The inverse half-life of $\onbb$-decay can be expressed as~\cite{Dekens:2020ttz} 
\begin{align}
\left(T_{1/2}^{0\nu}\right)^{-1} = g_A^4 G_{01} |\mathcal{A}_L(m_F)|^2\;,
\end{align}
where $g_A=1.271$, the phase space factor $G_{01}=1.5\times 10^{-14}~\text{year}^{-1}$ for $^{136}$Xe~\cite{Kotila:2012zza,Stoica:2013lka}. The amplitude 
\begin{align}
\label{eq:amp}
\mathcal{A}_L(m_F) &= -\dfrac{1}{2 m_e v} C_{\pi \pi L}^{(9)} \mathcal{M}_{PS}(m_F)\;,
\end{align}
where $m_e$ is the mass of the electron and $ \mathcal{M}_{PS}(m_F)$ is the nuclear matrix element (NME) that depends on the mass of $m_F$. Using the interpolation formulae developed in  Refs.~\cite{Dekens:2020ttz,Faessler:2014kka,Barea:2015zfa}, we find that  $\mathcal{M}_{PS}(m_F \geq 2\gev)=-0.44$ for $^{136}$Xe in the quasi-particle random phase approximation~\cite{Hyvarinen:2015bda}.
 
\begin{figure}[ht]
\centering
\includegraphics[width=0.7\linewidth]{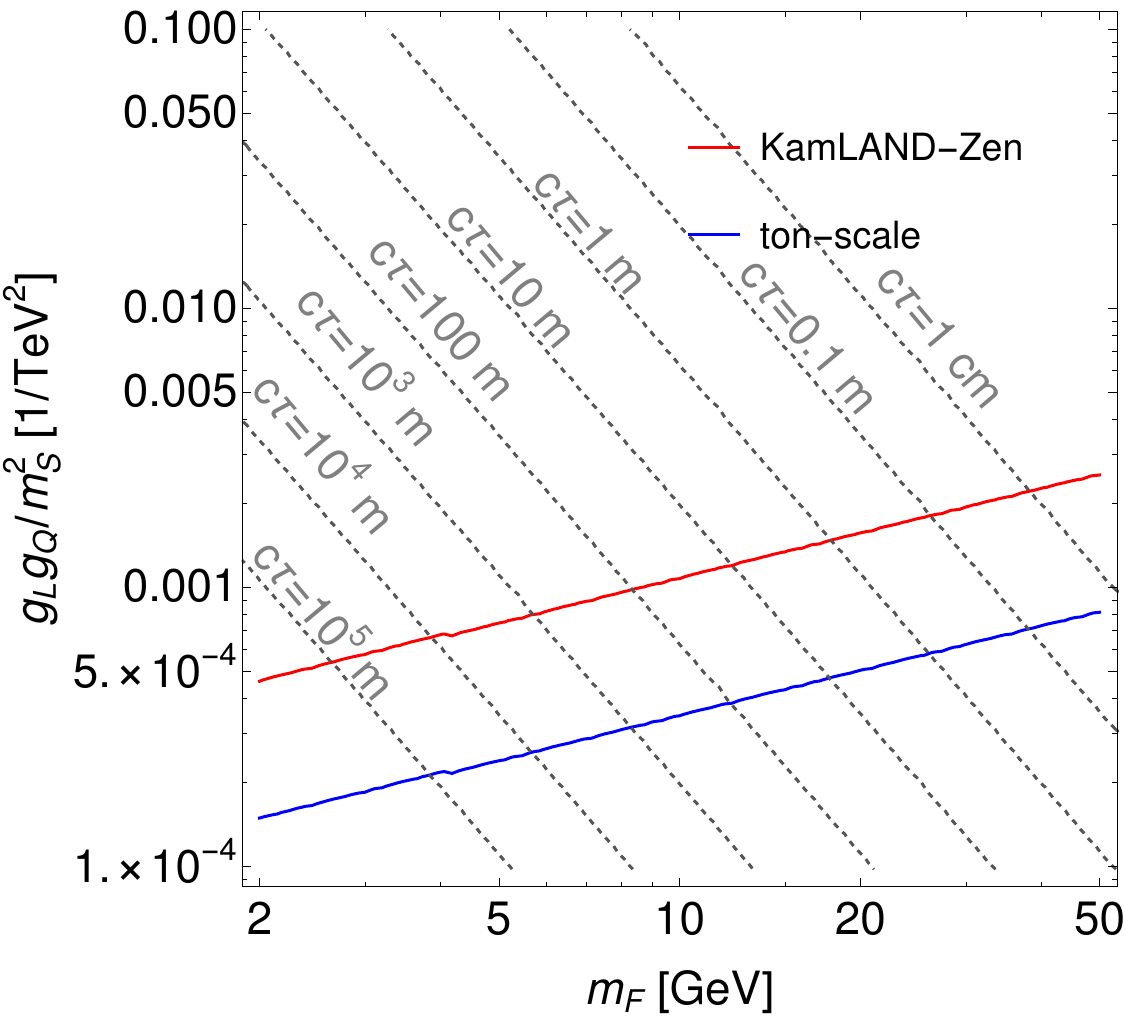}  
 \caption{$\onbb$-decay and proper decay length in case of light Majorana fermion $F$. The blue and red curves corresponds to the current limits and expected limits from KamLAND-Zen and ton-scale experiments. 
 The gray curves denote the proper decay lengths of $F$, which ranges from 1~cm to $10^{5}$~m.  }
 \label{fig:NLDBD}
\end{figure}

Figure~\ref{fig:NLDBD} illustrates the sensitivity of $\onbb$-decay to  $g_L g_Q/m_S^2$ and $m_F$. The red curve corresponds to the current limit from the KamLAND-Zen experiment~\cite{KamLAND-Zen:2016pfg}, $T_{1/2}^{0\nu}>1.07\times 10^{26}~\text{year}$ at 90\% confidence level (C.L.). The lower limit on the $\onbb$-decay half-life is expected to be improved by  2 orders of magnitude in a ton-scale detector,  reaching $T_{1/2}^{0\nu}>10^{28}~\text{year}$,  as depicted in blue color in Fig.~\ref{fig:NLDBD}. The regions between the blue and red curves are in the reach of ton-scale experiments~\cite{Kharusi:2018eqi,Abgrall:2017syy,CUPIDInterestGroup:2019inu,Paton:2019kgy,Chen:2016qcd,Adams:2020cye}. We can see that future $\onbb$-decay experiments can probe the light $m_F$ region $2\gev\leq m_F \leq 50\gev$ and put a strong bound on the effective coupling $g_L g_Q/m_S^2$~\footnote{Note that in Ref.~\cite{Peng:2015haa}, the notation $g_{\text{eff}}=g_L g_Q$ was introduced. }.

\section{Collider searches for a long-lived \tf{$F$}{F}}

\begin{figure}[ht]
\centering
\includegraphics[width=0.8\linewidth]{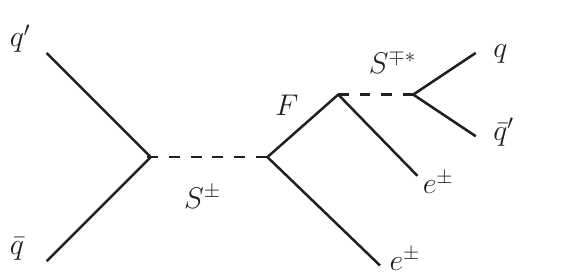}  
 \caption{The production and decay of Majorana fermion $F$ at the LHC with two same-sign electrons. Here, $(q,q^\prime)=(d,u), (u,d)$. The scalar $S^\pm$ is off shell in the decay of $F$ for $m_F<m_S$. }
 \label{fig:prod_decay}
\end{figure}

In this section, we discuss the LLP searches for the light Majorana fermion $F$ observed at the LHC with the main detectors or far detectors, which is complementary to the work done in  Ref.~\cite{Peng:2015haa} for the prompt decaying fermion. For $m_F< m_S$,  $F$ can be produced from $S^\pm \to F e^\pm$, and decay via an off-shell $S^\pm$ into $e^\pm q\bar{q}^\prime$ with $q\bar{q}^\prime$ being $d\bar{u}$ or $u\bar{d}$. The Feynman diagram for the production and decay of $F$ are shown in Fig.~\ref{fig:prod_decay}.  For a light $F$ and relatively small couplings, $F$ could be long-lived, which leads to the collider signature of a LLP with two same-sign electrons in the final state. Given the Majorana nature of $F$ and the couplings of 
$F$ to both $S^\pm$ and $S^0$, final states could include zero electron, one electron, or two electrons with either the same sign or the opposite signs.  In this study, we will focus on the same-sign dilepton process~\footnote{In Ref.~\cite{Aad:2015rba},  the ATLAS Collaboration has released a result searching for the displaced lepton pairs with opposite electric charges. If the charge identification of the electron from the decay of $F$ is feasible at the LHC main detector and far detectors, the LLP plus same sign dilepton signal could provide unambiguous evidence for LNV interactions.}  as depicted in Fig.~\ref{fig:prod_decay}.    The inclusion of opposite-sign dilepton and single-lepton processes could further enhance the sensitivity.

Previous studies considering a long-lived  heavy neutral lepton $N$, which may be a Majorana fermion, have focused on the processes: (a) $pp\to W^\pm$, $W^\pm \to \ell^\pm N$ via a SM $W$ boson, and  (b) $pp\to W_R^\pm, W_R^\pm \to \ell^\pm N$ via a heavy $W_R$, in the context of heavy-light neutrino mixing~\cite{Cottin:2018nms} and left-right symmetric model~\cite{Cottin:2018kmq,Nemevsek:2018bbt,Cottin:2019drg}, respectively. Our paradigm is different in three aspects. First, the production and decay of $F$ depend on different combinations of $g_L$ and $g_Q$ so that {a} sizable {number of} signal events and a long decay length of $F$ can be achieved simultaneously.  Second, $F$ is produced in the decay of a heavy resonance, yielding a much larger boost factor than process (a). Third, the couplings $g_L$ and $g_Q$ can be much smaller than the SM gauge couplings used in the process (b), leading to a longer decay length of $F$.  The decay length of $N$ in processes (a) and (b) is typically less than $\mathcal{O}(1)$~m~\cite{Cottin:2018kmq,Nemevsek:2018bbt}, while in our case, the decay length of $F$ can be orders of magnitude larger.

The total cross section $\sigma_{eF}$ of the processes $pp\to e^+ F$ and $pp\to e^- F$ can be  expressed as
\begin{align}
\sigma_{eF} = \sum_{i,j}\int \left(\dfrac{d \hat{s}}{\hat{s}} \right) \left( \dfrac{d\mathcal{L}_{ij}}{d\hat{s}} \right) \big(\hat{s} \hat{\sigma}_{ij}\big)\;,
\end{align}
where the differential parton-parton luminosity for partons $i$ and $j$~\cite{Campbell:2006wx,Clark:2016jgm} is
\begin{align}
\dfrac{d\mathcal{L}_{ij}}{d\hat{s}} = \dfrac{1}{s}\int_{\tau}^{1}\dfrac{dx}{x} &\Big[f_i (x,\sqrt{\hat{s}}) f_j (\dfrac{\tau}{x},\sqrt{\hat{s}}) \nn\\
&+ f_i (x,\sqrt{\hat{s}}) f_j (\dfrac{\tau}{x},\sqrt{\hat{s}}) \Big]\;
\end{align}
with $\tau=\hat{s}/s$ and $f_{i,j}(x,\sqrt{\hat{s}})$ is the parton distribution functions (PDFs) of partons $i,j$ evaluated at the momentum fraction $x$ and fractorization scale $\sqrt{\hat{s}}$. The partonic cross section $\hat{\sigma}_{ij}$ is non-vanishing only for $(i,j)=(u,\bar{d})$, $(d,\bar{u})$,
\begin{align}
\hat{\sigma}_{u\bar{d}} &= \hat{\sigma}_{d\bar{u}}= \hat{\sigma}_{S} {\rm Br}(S\to F e)\;,\\
\hat{\sigma}_{S} &= \dfrac{1}{24\hat{s}} \left( g_Q^2 m_S^2\right) (2\pi) \delta (\hat{s}-m_S^2)\;,
\end{align}
 where $\hat{\sigma}_{S}$ and ${\rm Br}(S\to F e)$  denote the production cross section of $u\bar{d}\to S^+$ and decay branching ratio of $S^+\to F e^+$, respectively. 
The convolution of parton distribution functions is evaluated using \texttt{ManeParse} package~\cite{Clark:2016jgm} with the PDF set NNPDF3.0NLO~\cite{NNPDF:2014otw}. We have checked that the cross section agrees well with that obtained using \texttt{MadGraph5\_aMC@NLO}~\cite{Alwall:2014hca}. 

Given the interactions in Eq.~\eqref{eq:Lag}, a light $F$ can only decay to three-body final states via an off-shell $S^\pm$ or $S^0$ if $m_F < m_S$:
\begin{align}
F\to e^+ \bar{u}d\;,\ e^-\bar{d}u\;,\ \nu_e d\bar{d}\;,\ \bar{\nu}_e d\bar{d}\;.
\end{align}
Following Refs.~\cite{Bondarenko:2018ptm,deVries:2020qns},  the inclusive hadronic decay width of $F$ is given by  
\begin{align}
\Gamma_{\rm tot} = (1+\Delta_{\text{QCD}})\ \left( \Gamma_{eud} + \Gamma_{\nu dd}\right)\;,
\end{align} 
where
$\Gamma_{eud} $ denotes the partial widths into $e^+ \bar{u}d$ and $e^- \bar{d}u$, and $\Gamma_{\nu dd}$ is the sum of partial widths into $\nu_e d \bar{d}$ and $\bar{\nu}_e d \bar{d}$. For $m_F \ll m_S$, one has
~\footnote{Strictly speaking, the expression of $\Delta_{\text{QCD}}$ in Eq.~\eqref{eq:Delta_QCD} is only valid for the purely left-handed charged current interactions. We will neglect this difference given the relative small size of $\Delta_{\text{QCD}}$.} 
\begin{align}
&\Gamma_{eud} = 2\ \Gamma_{\nu dd}= \dfrac{m_F^5 g_L^2 g_Q^2}{1024\pi^3 m_S^4 } \;,\\
\label{eq:Delta_QCD}
&\Delta_{\text{QCD}} = \dfrac{\hat\alpha_S}{\pi} + 5.2\left( \dfrac{\hat\alpha_S}{\pi} \right)^2 + 26.4\left( \dfrac{\hat\alpha_S}{\pi} \right)^3
\end{align}
with $\hat\alpha_S \equiv \alpha_S(m_F)$. The correction $\Delta_{\text{QCD}}$ varies from 16\% to 5\% for $2\gev\leq m_F\leq 50\gev$. The decay branching ratio of $F\to e^+ \bar{u} d$ and $F\to e^- u \bar{d}$ via an off-shell $S^\pm$ is $\text{Br}_{ejj} \equiv \text{Br}(F\to e^+ \bar{u} d) = \text{Br}(F\to e^- u \bar{d}) =1/3$ with $j$ denoting inclusively the first-generation quarks or antiquarks. 

In Fig.~\ref{fig:NLDBD} we also show the proper decay length of $F$, $c\tau = \hslash c/\Gamma_{\rm tot}$, which varies from $1~\text{cm}$ to $10^5~\text{m}$  for  $2\gev\leq m_F\leq 50\gev$.   The $\onbb$-decay half-life for $3\times 10^{-4}\tev^{-2}\leq g_L g_Q/m_S^2 \leq 5\times 10^{-3} \tev^{-2}$  is within the reach of future ton-scale experiments, enabling the interplay of $\onbb$-decay and LLP searches.

In the following, we will investigate sensitivities to a long-lived particle $F$ at the LHC main detectors ATLAS/CMS and a proposed far detector MATHUSLA~\cite{Curtin:2018mvb}  for the signal process in Fig.~\ref{fig:prod_decay}. The observed number of events $N_{\text{obs}}^{\text{detector}}$ from the decay of long-lived $F$ can be estimated with the following formula~\cite{Bauer:2018uxu} 
\begin{align}
\label{eq:observed_events}
N_{\text{obs}}^{\text{detector}}
&= \sigma_{eF}\ \text{Br}_{ejj}\ \mathcal{L}\ \epsilon_{\text{LLP}}^{\text{detector}}\ \epsilon_{\text{prompt}}^{\text{detector}}\ \mathcal{P}_{\text{decay}} \;,
\end{align}
where 
\begin{align}
\label{eq:prob_eff}
\mathcal{P}_{\text{decay}} \equiv\dfrac{1}{\sigma_{eF}} \int_{\Delta \Omega} d\Omega \dfrac{d\sigma_{eF}}{d\Omega} \int_{L_1}^{L_2} dL \dfrac{1}{d_\perp} e^{-L/d_\perp }\;.
\end{align}
 Here, $\sigma_{eF}\ (d\sigma_{eF}/d\Omega)$ denotes the total (differential) cross section of $pp\to S^\pm$, $S^\pm \to e^\pm F$ in the lab frame, and $d \Omega = d\cos\theta d\phi$ with $\phi$ $(\theta)$ being the azimuthal (polar) angle of $F$ with respect to the beam axis. $\Delta\Omega$  is the solid angle coverage of the LLP detector. $L_1$ and $L_2$ $(L_1<L_2)$ are the distances of the LLP detector from the LHC interaction point. The transverse decay length $d_\perp = d \sin\theta$, where $d$ is the boosted decay length of $F$. Since $F$ is produced from the decay of the charged scalar $S^\pm$, it is highly boosted with the boost factor $b=\beta\gamma=m_S/(2m_F)$, and the boosted decay length of $F$ in the rest frame of $S$ (effectively the lab frame) is $d=bc\tau $. In Eq.~\eqref{eq:observed_events}, ``detector'' = ``LHC'', ``MATH'' refer to   the LHC main detectors ATLAS/CMS, and MATHUSLA, respectively. $\epsilon_{\text{LLP}}^{\text{detector}}$ is the efficiency with LLP cuts, while $\epsilon_{\text{prompt}}^{\text{detector}}$ denotes the efficiency of prompt cuts.  We consider the integrated luminosity  $\mathcal{L}=3\abi$ at the HL-LHC. 

Given the distance of the MATHUSLA detector to the interaction point and the relative small solid angle span of the detector at the MATHUSLA, it is convenient to approximate Eq.~\eqref{eq:observed_events} as~\cite{Curtin:2018mvb}:
\begin{align}
N_{\text{obs}}^{\text{MATH}} &= \sigma_{eF}\  \text{Br}_{ejj}  \ \mathcal{L} \ \epsilon_{\text{LLP}}^{\text{MATH}}\ \epsilon_{\text{geometric}}\ P_{\text{decay}}\;,
\end{align}
with the geometric acceptance $\epsilon_{\text{geometric}} =0.05$ describing the fraction of LLPs traversing the MATHUSLA detector. The probability for $F$ decaying inside the detector is
\begin{align}
\label{eq:prob}
P_{\text{decay}}(d; L_1,L_2) 
=  e^{-L_1/d}-e^{-L_2/d}\;.
\end{align}
For the MATHUSLA detector, $L_1=200$~m, $L_2=230$~m, and the detection efficiency per decay $\epsilon_{\text{LLP}}^{\text{MATH}}$  of LLPs within the detector volume is approximately equal to 1 for hadronic decays and $0.5\sim 1$ for leptonic 2-body decays~\cite{Curtin:2018mvb}. Although the decay products of $F$ could be collimated, we have checked that their relative polar angles $\Delta \theta$ are always larger than 0.01 (a reference value suggested in Ref.~\cite{Curtin:2018mvb}) for the masses of $S^\pm $ and $F$ of interest in this study. Therefore, it is reasonable to assume that the decay productions can be isolated and the efficiency $\epsilon_{\text{LLP}}^{\text{MATH}}=1$.

At the LHC main detectors ATLAS/CMS, the steradian coverage $\Delta\Omega=4\pi$ in the solid angle~\cite{Aad:2015uaa}, we will use the following expression to approximate  Eq.~\eqref{eq:observed_events}, assuming isotropic angular distribution~\cite{Antusch:2018svb}:   
\begin{align}
\label{eq:obs_LHC}
N_{\text{obs}}^{\text{LHC}} = \sigma_{eF}\  \text{Br}_{ejj} \  \mathcal{L}\ \epsilon_{\text{LLP}}^{\text{LHC}}\ \epsilon_{\text{prompt}}^{\text{LHC}}\ P_{\text{decay}}\;,
\end{align}
where $L_1=2~\text{cm}$ and $L_2=1.1~\text{m}$~\cite{Aaboud:2017iio,Aad:2015rba,Aad:2019xav,Davoudiasl:2021haa} for the inner detector of ATLAS/CMS. Note that reach could be enhanced if muon chamber is also used for LLP search~\cite{Aad:2015uaa}. 
We will assume that $\epsilon_{\text{LLP}}^{\text{LHC}}\ \epsilon_{\text{prompt}}^{\text{LHC}} = 0.01$~\cite{Cottin:2018kmq} to approximate the trigger, identification, and cut efficiency of detecting LLPs at the LHC main detectors\footnote{Note that as mentioned in Ref.~\cite{Curtin:2018mvb}, the typical value of $\epsilon_{\text{LLP}}^{\text{LHC}}$ is about 0.1 for displaced vertex (DV) searches in ATLAS tracker and 0.5 for displaced jet searches in tracker.  The efficiency is expected to be reduced if the decay products of light $F$ are collimated. In the search of heavy neutral lepton $N$ in the left-right symmetric model~\cite{Cottin:2018kmq}, it was found that the $\epsilon_{\text{LLP}}^{\text{LHC}}\ \epsilon_{\text{prompt}}^{\text{LHC}}= 6\times 10^{-4}$ using the ATLAS standard DV cuts~\cite{Aaboud:2017iio} for $W_R$ mass $m_{W_R}=4\tev$ and $m_{N}=20\gev$. However, as pointed out in Ref.~\cite{Cottin:2018kmq}, this efficiency  could be improved to  $\epsilon_{\text{LLP}}^{\text{LHC}}\ \epsilon_{\text{prompt}}^{\text{LHC}}= 0.026$ by loosening the requirements of the DV invariant mass and the number of tracks.    We have checked that the angular separation $\Delta R$ between any two objects in the decay of $N$ has the peak around 0.02~\cite{Mitra:2016kov}, while $\Delta R$ between any two objects in the decay of $F$ in our case tends to be 0.1 for $m_S=2\tev$ and even larger for a smaller $m_S$,  which makes the final states easier to be reconstructed. Instead of reanalyzing the DV searches~\cite{Aaboud:2017iio} as in Ref.~\cite{Cottin:2018kmq}, we will assume a medium value $\epsilon_{\text{LLP}}^{\text{LHC}}\ \epsilon_{\text{prompt}}^{\text{LHC}}=0.01$, which should be realistic for the LHC LLP searches with ATLAS/CMS.
}.
Finally, it is worthwhile to note that the approximate expression in Eq.~\eqref{eq:obs_LHC} is valid under the assumption that $S$ decays isotropically and the decay length is used in the decay probability function.
Deviation from  this assumption might introduce another factor of 2 uncertainty~\footnote{For instance, for $m_F=50\gev$ and $g_{\text{eff}}=10^{-3} $, the ratio $P_{\text{decay}}/\mathcal{P}_{\text{decay}}$ is 1.19 and 0.67 for $m_S=1\tev$ and $2\tev$, respectively.}, which is incorporated into the assumption of $\epsilon_{\text{LLP}}^{\text{LHC}}\ \epsilon_{\text{prompt}}^{\text{LHC}}=0.01$. 

Existing LHC searches  also constrain the model parameter space.
Ref.~\cite{Bordone:2021cca} performed a combination searches for dijet resonances within the mass range $450\gev$ to $5\tev$~\cite{ATLAS:2018qto,CMS:2018mgb,ATLAS:2019fgd,Sirunyan:2019vgj}, in which the difference in the acceptance for resonances of a different spin was found to be small and can be neglected. We have verified it by comparing its constraint for a scalar against the CMS Collaboration~\cite{Sirunyan:2019vgj} for the scalar mass above 1.8~TeV, and find a good agreement.
Hence, we will follow the procedure in Ref.~\cite{Bordone:2021cca} to  obtain the combined constraints on the sum of $\sigma(pp\to S^\pm )\times{\rm Br}(S^\pm \to j j)$ and $\sigma(pp\to S^0 )\times{\rm Br}(S^0 \to j j)$ in the mass region of $450\gev\leq m_S \leq 5\tev$. We also investigated the current searches~\cite{CMS:2018jxx} for Majorana neutrino with same-sign dilepton signature. It is relevant if the lifetime of $F$ in our case is small and decays inside the detector, which requires large couplings $g_{L,Q}$ and/or large mass $m_F$.  We have checked that such constraint on the parameter space that we are interested in is weak.

Besides, the missing energy searches can also constrain the parameter space although they cannot test whether the lepton number is violated or not. For Majorana fermion $F$ decaying outside of the LHC main detectors, the result of searching for a tau slepton decaying into electron and missing energy~\cite{CMS:2021pmn} can be reinterpreted directly to obtain an upper bound on $\sigma(pp\to S^\pm)\times{\rm Br}(S^\pm \to e^\pm F)\exp(-L_2/d)$~\cite{Dev:2019hho}. Here $\exp(-L_2/d)$ characterizes the probability of $F$ decaying outside of the CMS detector with detector cross section radius of $L_2=7.5$~m~\cite{CMSbook}. There also exist direct searches for pair production of sleptons, which decay into the lighest neutralinos in the context of R-parity conserving SUSY models at the LHC with the integrated luminosity of $137\fbi$. It excludes slepton mass below $700\gev$~\cite{CMS:2020bfa} combining the electron and muon decay channels and different chiralities of sleptons. This might constrain our model via $pp \to S^+ S^-,S^\pm \to e^\pm F$ with $F$ decaying outside the detector.  However, the constraint is expected to be relatively weak and therefore not considered hereafter.  Finally, we consider one-loop contribution to the electron anomalous magnetic moment, $\Delta a_e = - g_e^2/(16\pi^2) m_e^2/m_S^2 f(m_F^2/m_S^2)$ with the loop function $f(x) \simeq 1/6-x/3 $ for $x\ll 1$~\cite{Liu:2021mhn}. We find that $\Delta a_e$ is negative and its magnitude is below $2.8\times 10^{-16} (1\tev/m_S)^2$ for $g_e\leq 1$, hence there is no constraint from the experimental measurements of the electron anomalous magnetic moment~\cite{Hanneke:2008tm,Hanneke:2010au,2018Science}.

\begin{figure}[ht]
\centering
\includegraphics[width=0.48\linewidth]{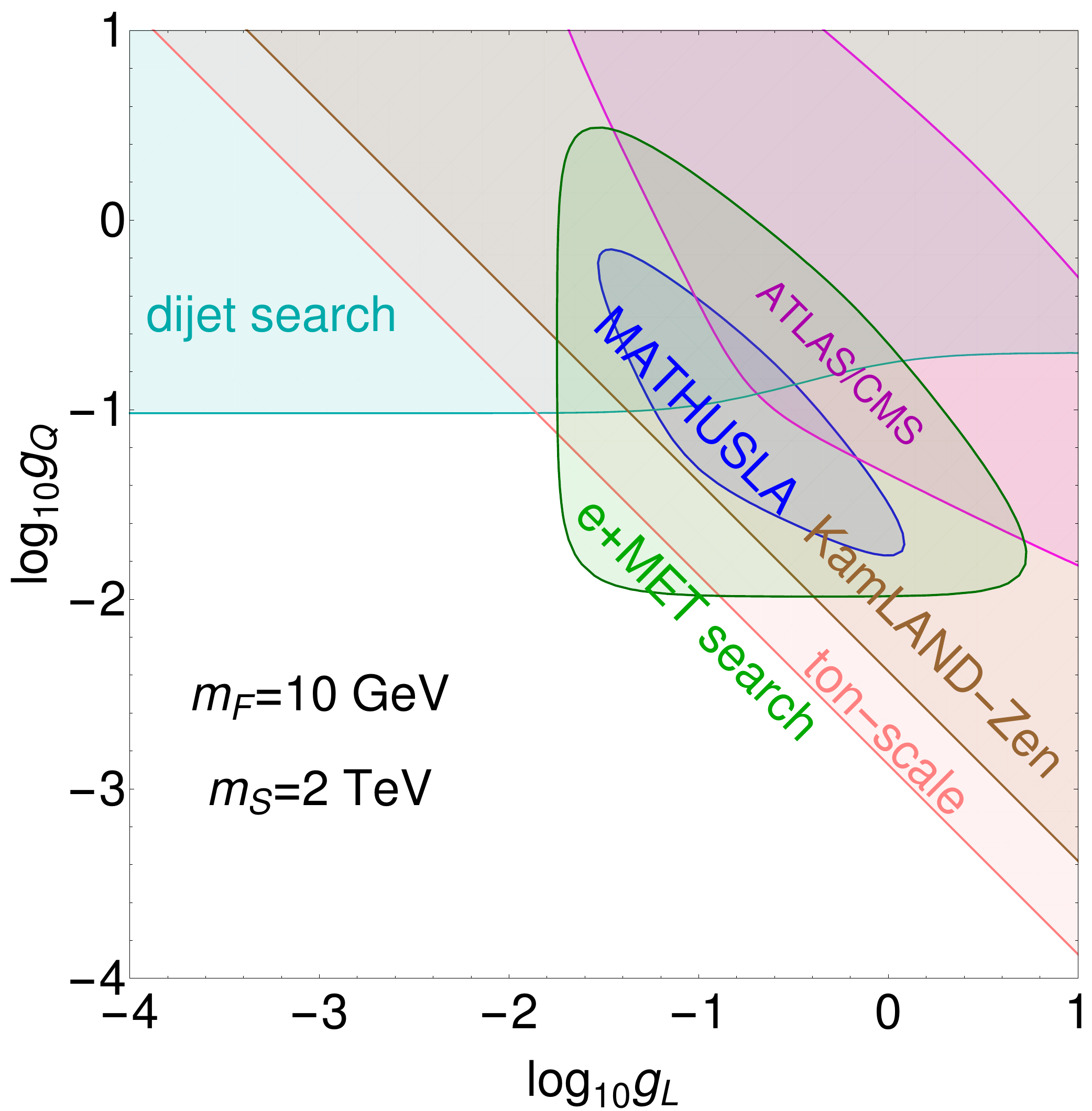}
\includegraphics[width=0.48\linewidth]{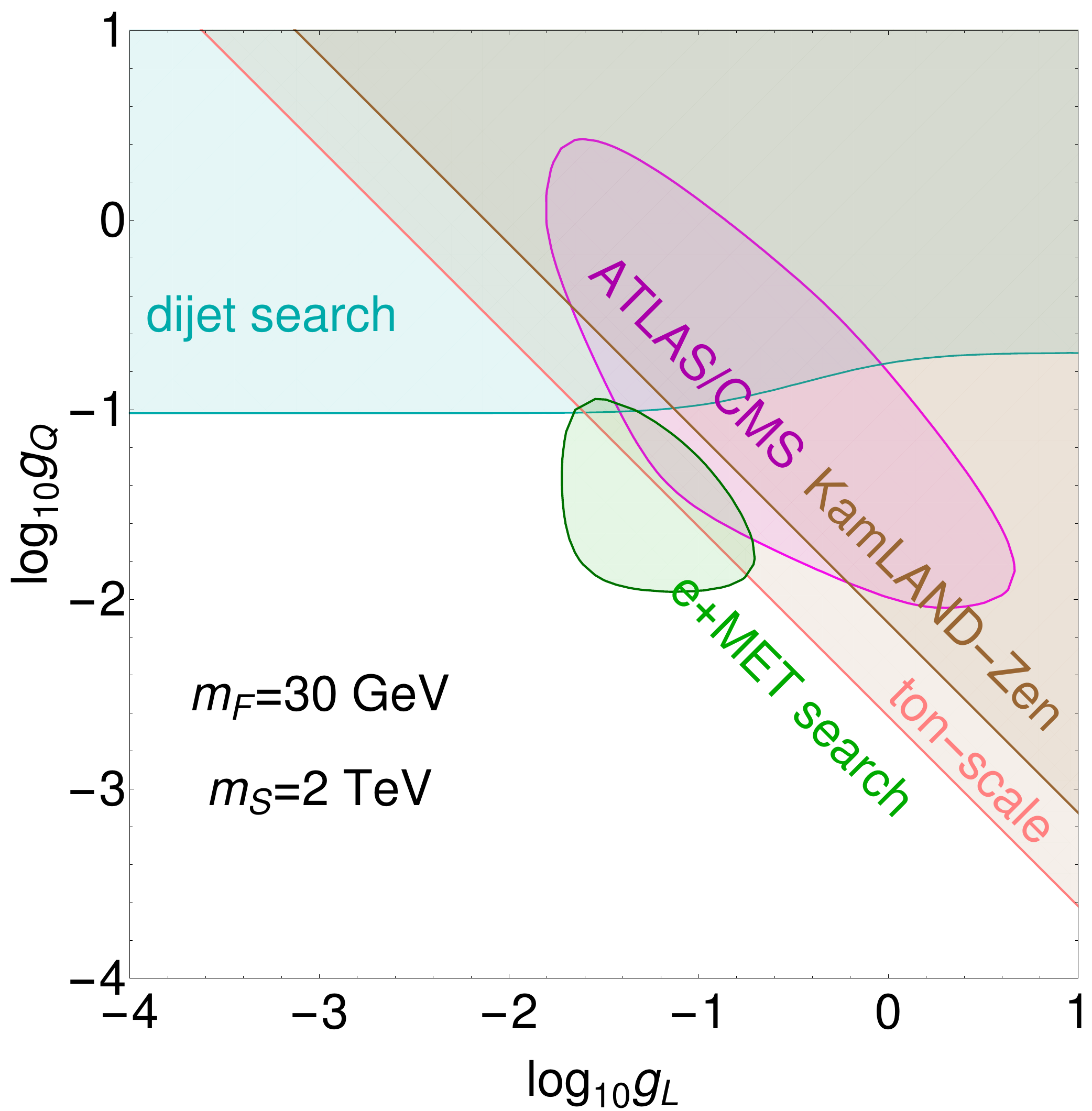}
\caption{The combined sensitivities in $\onbb$-decay and LLP searches for $m_F=10,30\gev$ for $m_S=2\tev$. The shaded regions denote the exclusion limits by the KamLAND-Zen (brown), future ton-scale (pink) $\onbb$-decay experiments, and LLP searches at ATLAS/CMS (magenta), and MATHUSLA (blue). The current prompt searches for a resonance in the dijet final state~\cite{ATLAS:2019fgd} (cyan), and the final state of electron and missing energy~\cite{CMS:2021pmn,CMS:Oh} (green) at the 13~TeV LHC with the integrated luminosities of $139\fbi$ and $137\fbi$ respectively are also considered.}
\label{fig:MATH_LHC_combined}
\end{figure}

\begin{figure}[ht]
\centering
\includegraphics[width=0.48\linewidth]{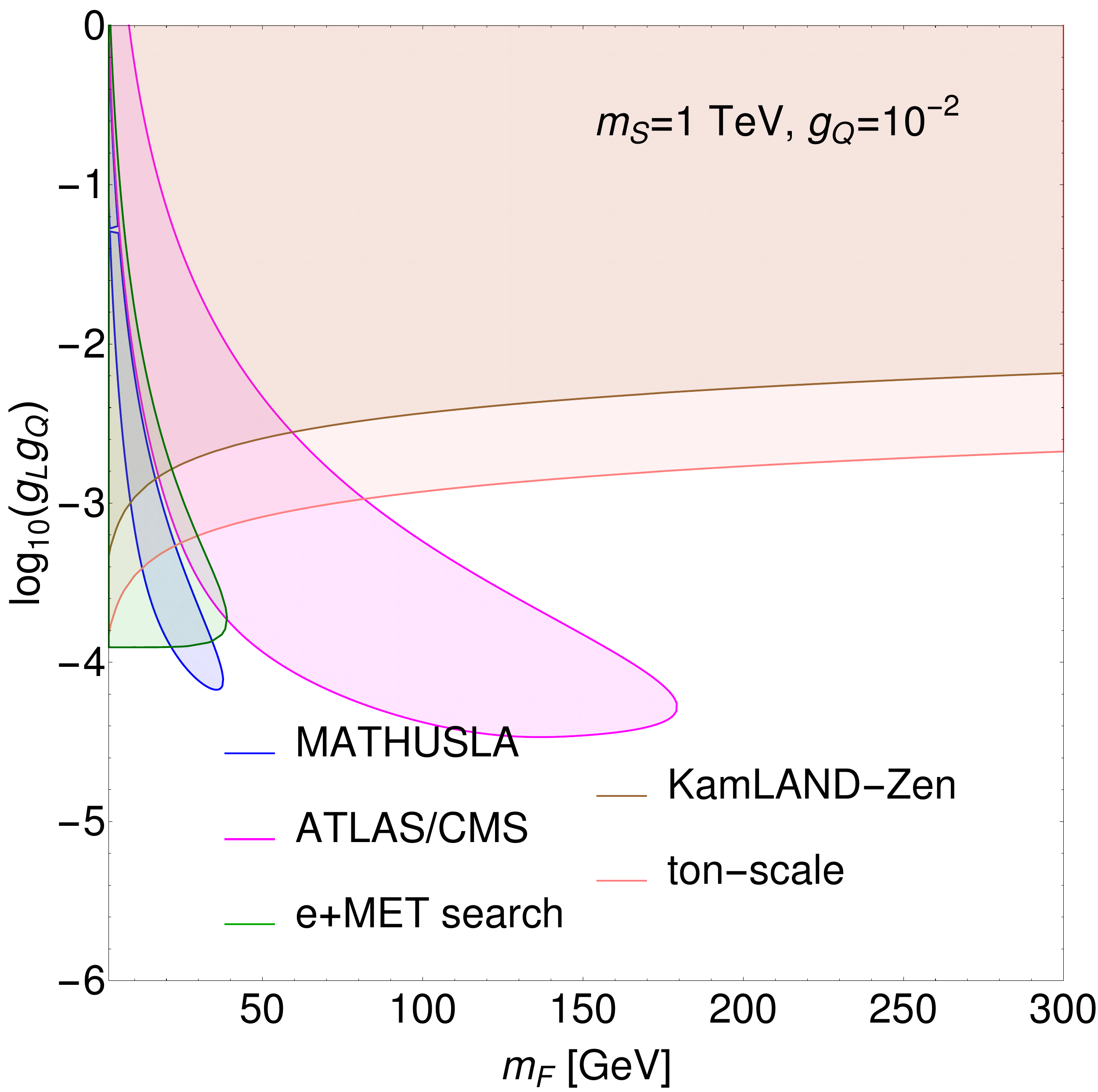} 
\includegraphics[width=0.48\linewidth]{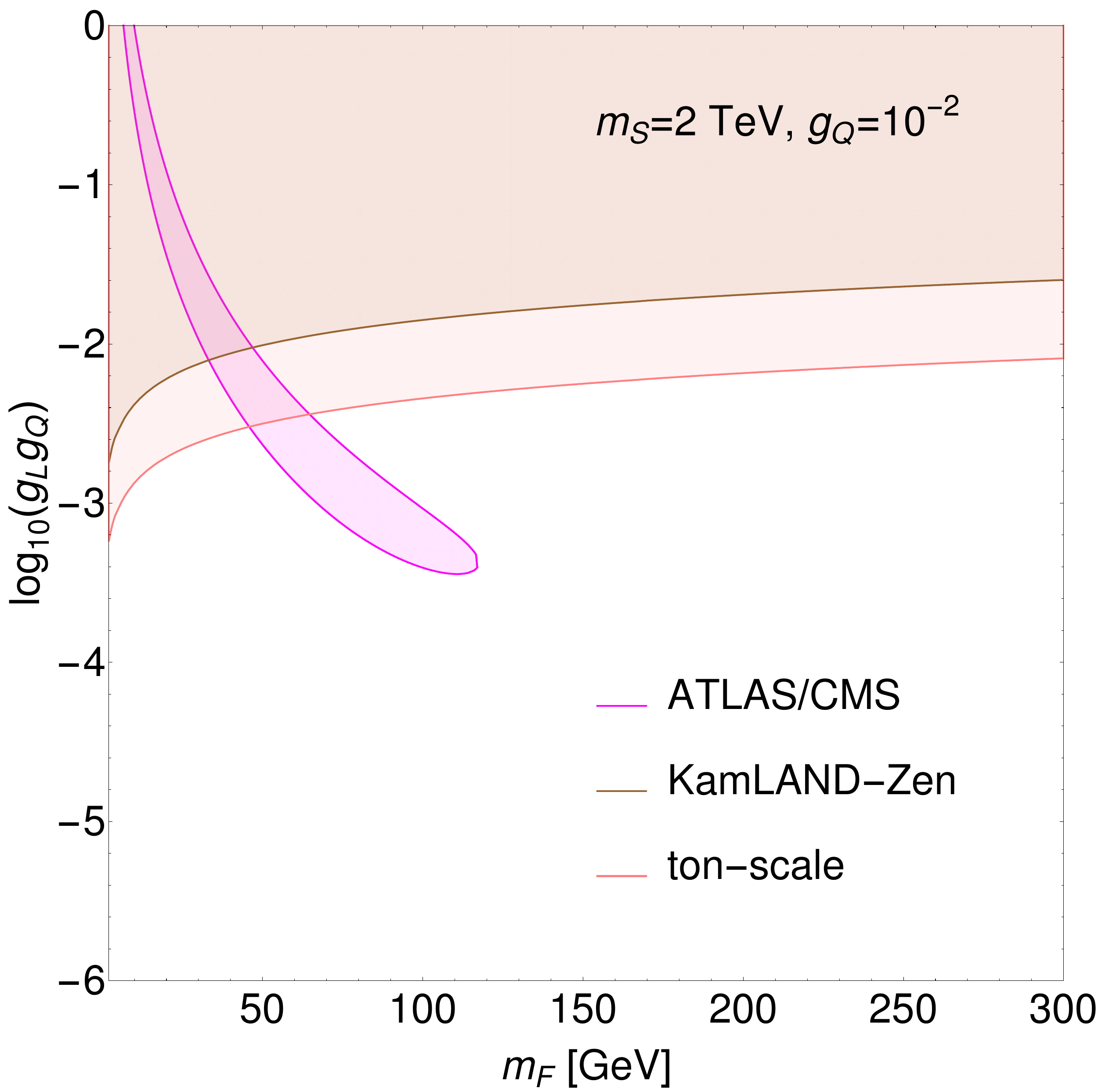} 
\caption{Sensitivities in $\onbb$-decay and LLP searches in the plane of $m_F$ and $g_L g_Q$ for  $g_Q=10^{-2}$ and $m_S=1\tev$ (left panel) or $m_S=2\tev$ (right panel). The shaded regions are excluded by the KamLAND-Zen (brown), future ton-scale (pink) $\onbb$-decay experiments, and LLP searches at ATLAS/CMS (magenta), MATHUSLA (blue). The green region is excluded by the electron and missing energy search~\cite{CMS:2021pmn} (green). There is no constraint from the dijet searches~\cite{ATLAS:2018qto,CMS:2018mgb,ATLAS:2019fgd,Sirunyan:2019vgj} with $g_Q=10^{-2}$.}
\label{fig:mF_geff}
\end{figure}

\begin{figure}[ht]
\centering
\includegraphics[width=0.48\linewidth]{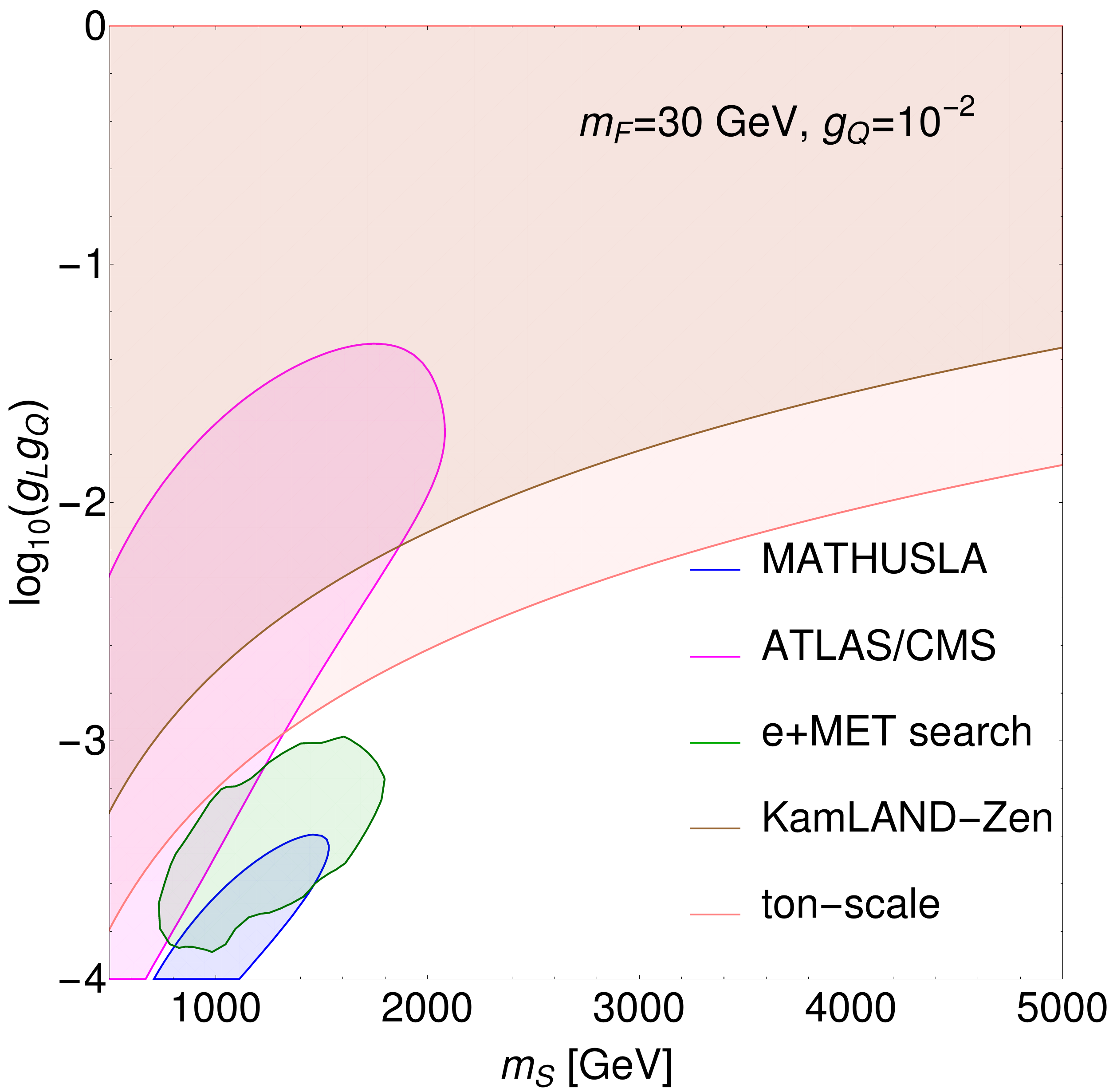}  
\includegraphics[width=0.48\linewidth]{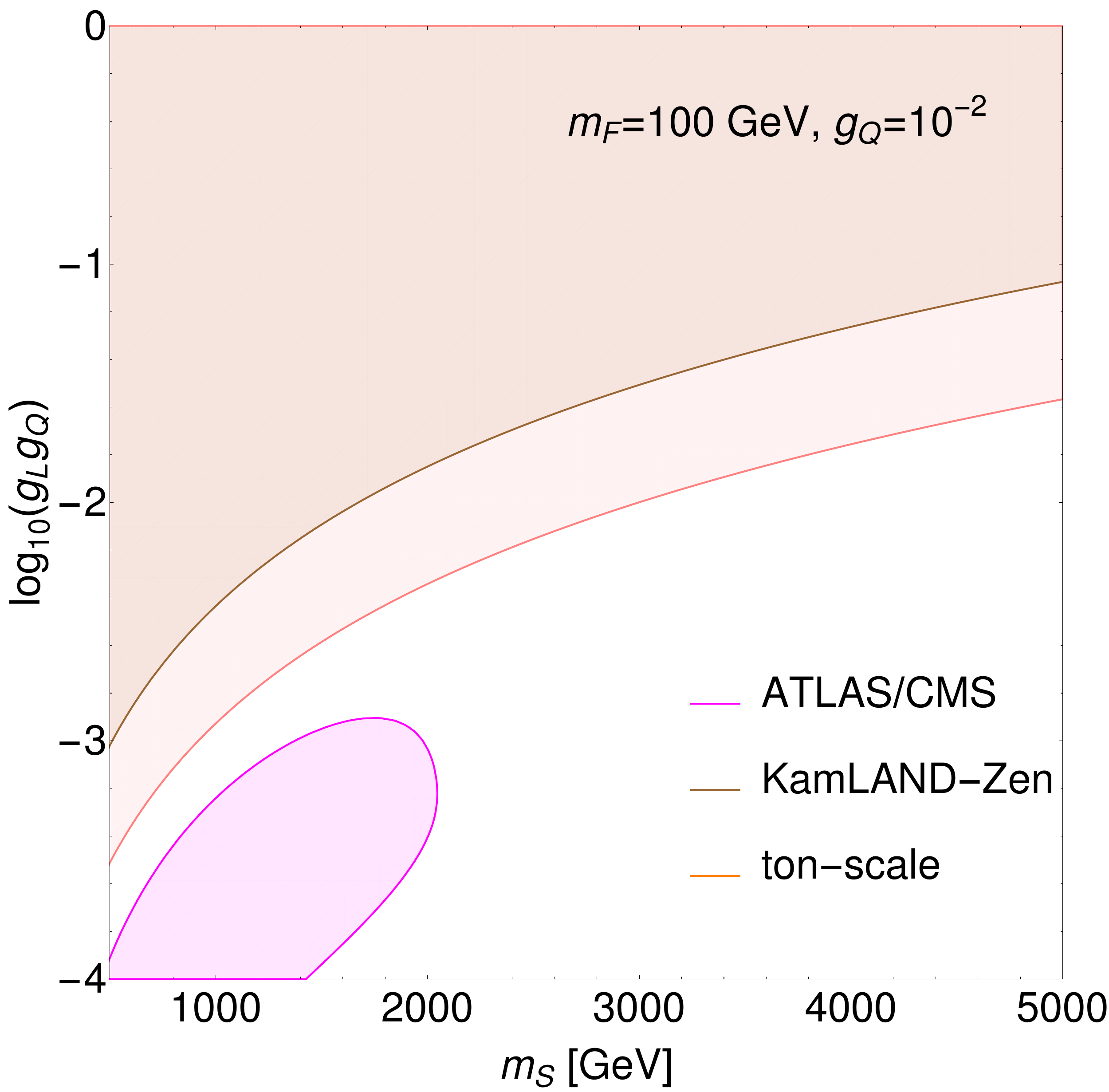}  
\caption{Sensitivities in $\onbb$-decay and LLP searches in the plane of $m_S$ and $g_L g_Q$ for $g_Q=10^{-2}$ and $m_F=30\gev$ (left panel) or $m_F=100\gev$ (right panel). The shaded regions are excluded by the KamLAND-Zen (brown), future ton-scale (pink) $\onbb$-decay experiments, and LLP searches at ATLAS/CMS (magenta), MATHUSLA (blue). The green region is excluded by the electron and missing energy search~\cite{CMS:2021pmn} (green). There is no constraint from the dijet searches~\cite{ATLAS:2018qto,CMS:2018mgb,ATLAS:2019fgd,Sirunyan:2019vgj} with $g_Q=10^{-2}$.}
\label{fig:mS_geff}
\end{figure}

The 95\% C.L. exclusion limits of ATLAS/CMS and MATHUSLA at the HL-LHC are obtained by requiring $N_{\text{obs}}=3$ assuming zero background~\cite{Junk:1999kv,Bhattiprolu:2020mwi},
which are shown  in magenta and blue, respectively,  in Fig.~\ref{fig:MATH_LHC_combined} $-$ Fig.~\ref{fig:mS_geff}. The current constraints from dijet search~\cite{Sirunyan:2019vgj} and electron and missing energy search~\cite{CMS:2021pmn, CMS:Oh} at the LHC are shown in cyan and green, respectively.  
The KamLAND-Zen and future ton-scale $\onbb$-decay experiments set a lower bound on the half-life of $\onbb$-decay, which excludes the brown and pink regions, respectively.
In Fig.~\ref{fig:MATH_LHC_combined}, the results are shown in the plane of the couplings $g_L$ and $g_Q$ for $m_F=10,30\gev$.
For light $F$ in the left panel, the LLP searches cannot compete with the $\onbb$-decay searches since the decay length of $F$ is too long, as seen in Fig.~\ref{fig:NLDBD}. The decay probability inside the detectors is highly suppressed, except in the large coupling region of the upper-right corner of the left panel. On the contrary, the electron and missing energy search have excluded regions of $g_{L,Q}$ around $10^{-2} - 1$,  even extending beyond the reach of ton-scale experiments.  

For larger $m_F$ in the right panel,   the LLP searches at ATLAS/CMS are complementary to the $\onbb$-decay searches in ton-scale experiments and can test the parameter space of $g_L$ or $g_Q$ larger than about $10^{-2}$, in which region the electron and missing energy search lose sensitivity due to short lifetime of $F$. 
There is no contour giving three signal events for MATHUSLA in the right panel since the sensitivity of LLP searches drops dramatically with the increase of $m_F$: $\Gamma_{\rm tot}\propto m_F^5$.

From the discussions in Sec.~\ref{sec:model_DBD}, the half-life of $\onbb$-decay is proportional to $[g_L^2 g_Q^2/(m_S^4 m_F)]^{-2}$, which implies that the sensitivity of $\onbb$-decay searches decreases for both larger $m_F$ and smaller $g_L g_Q/m_S^2$. On the other hand, in the LLP searches the lifetime of $F$ 
becomes smaller for larger $m_F$, but gets larger for smaller $g_L g_Q/m_S^2$. Therefore, it is expected that the LLP searches might uniquely probe the region of small $g_Lg_Q/m_S^2$. To see it more clearly, we plot the 95\% C.L. exclusion contours in the plane of $m_F$ and $g_L g_Q$ for $g_Q=10^{-2}$ and $m_S=1,2\tev$ in  Fig.~\ref{fig:mF_geff}. In both panels, the $\onbb$-decay searches are more sensitive to $g_Lg_Q\gtrsim 10^{-2}$ for all values of $m_F<300$ GeV, while the LLP search at ATLAS/CMS can probe smaller $g_Lg_Q$ for $m_F<200$ GeV. The LLP search at MATHUSLA, on the other hand, is only sensitive to light $m_F<50$ GeV for $m_S=1$ TeV.

The collider reach also depends sensitively on $m_S$ since the production of $pp\rightarrow S^\pm$ decreases rapidly as $m_S$ increases. For instance, the cross sections for $m_S=0.5,1,2\tev$ scale as $1:0.08:0.0036$.
In Fig.~\ref{fig:mS_geff}, we show the sensitivities in the plane of $m_S$ and $g_L g_Q$ for $g_Q=10^{-2}$ and $m_F=30,100\gev$. For $m_F=30\gev$, the LLP searches at ATLAS/CMS   mostly overlap with that of the $\onbb$-decay experiments.   The LLP searches at MATHUSLA, however, could extensively probe the region with $m_S\lesssim 1.8\tev$ with smaller $g_L g_Q$. For $m_F =100\gev$, the LLP searches at ATLAS/CMS and $\onbb$-decay are sensitive to different regions of $g_L g_Q$, proving a complementary test of lepton number violation.

\section{Conclusion}
This work has focused on the interplay between $\onbb$-decay and LLP searches as probes of TeV-scale lepton number violation. We have utilized a simplified model that generates the LO, long-range $\onbb$-decay amplitude and that adds to the Standard Model a minimal set of new particles and interactions -- a framework that one expects to hold in complete theories such as the RPV SUSY. In this context, our analysis demonstrates the complementarity between $\onbb$-decay and LLP searches with ATLAS/CMS and the proposed MATHUSLA detector as probes of the parameter space of
 lepton number violation. The LNV interactions may involve particles below the electroweak scale, giving positive $\onbb$-decay signals and a long decay length. 

These $\onbb$-decay and LLP searches have different dependence on the parameters, namely the couplings $g_L$, $g_Q$ and the masses $m_S$ and $m_F$. The $\onbb$-decay depends  on the combination $g_L^2 g_Q^2/(m_S^4 m_F)$. For the LLP searches, the production cross section of $pp\to S^\pm \to e^\pm F$ gets increased for larger $g_Q$, smaller $m_S$, and larger $g_L/g_Q$.  
The probability of $F$ traveling long enough and decaying inside the LLP detectors is sensitive to the decay length of $F$.  Small $g_L g_Q$ and small $m_F$ are typically needed to have a long enough decay lifetime.

By comparing the reaches of the current and future $\onbb$-decay experiments with the LLP searches at the LHC main detectors and MATHUSLA, while taking into account the relevant current LHC searches of prompt decays, we found that the $\onbb$-decay experiments can set the upper limits on $g_L g_Q$, ranging from $10^{-4}$ to $10^{-1}$ depending on the masses $m_S$ and $m_F$. The LLP searches at the LHC main detectors are sensitive to the regions of $m_F\lesssim 200\gev$, $m_S\lesssim 2\tev$, and small $g_L g_Q$ that are beyond the $\onbb$-decay reach. The MATHUSLA detector can further test the region with smaller $g_L g_Q$ and $m_F$.

Similar features as found in this work are expected to appear in other models with low-scale lepton number violation that induce the LO, long-range $\onbb$-decay amplitude, although the sensitivities to the parameter space could be quite different. With the upcoming running of the LHC and the proposed future experiment MATHUSLA, a better understanding of 
the  viable parameter space for LNV interactions
can be achieved in combination with the current and future $\onbb$-decay experiments.

\begin{acknowledgments}
GL would like to thank Jordy de Vries for many helpful discussions. Discussions with Lingfeng Li, Zhen Liu, Sebastian Urrutia-Quiroga,  Liantao Wang, Guanghui Zhou, are also thankfully acknowledged. 
SS is supported by the Department of Energy under Grant No.~DE-FG02-13ER41976/DE-SC0009913. JCV was supported in part under the U.S. Department of Energy contracts DE-SC0015376. GL, MJRM, and JCV were supported in part under U.S. Department of Energy contract DE-SC0011095.  MJRM was also supported in part under National Science Foundation of China grant No. 19Z103010239.
\end{acknowledgments}

\bibliographystyle{apsrev4-1}

\bibliography{reference}

%%%%%%%%%%%%%%%%%%%%%%%%%%%%%%%%%%%%%%%%%%%%%%%%%%%%%%

\end{document}